\documentclass[a4wide]{article}
\usepackage[latin1]{inputenc}
\usepackage[english]{babel}
\usepackage{times}
\usepackage{amsmath}
\usepackage{amssymb}
\usepackage{amsthm}
\usepackage{a4wide}
\usepackage{graphicx}

\begin{document}
\title{Following red blood cells in a pulmonary capillary}
\author{Benjamin Mauroy\\ {\small Laboratoire MSC, Universit\'e Paris 7 / CNRS,}\\ {\small 10 rue Alice Domon et L\'eonie Duquet, 75013 Paris.}}

\maketitle 

\begin{abstract}
The red blood cells or erythrocytes are biconcave shaped cells and consist mostly in a membrane delimiting a cytosol with a high concentration in hemoglobin. This membrane is highly deformable and allows the cells to go through narrow passages like the capillaries which diameters can be much smaller than red blood cells one. They carry oxygen thanks to hemoglobin, a complex molecule that have very high affinity for oxygen. The capacity of erythrocytes to load and unload oxygen is thus a determinant factor in their efficacy. In this paper, we will focus on the pulmonary capillary where red blood cells capture oxygen. We propose a camera method in order to numerically study the behavior of the red blood cell along a whole capillary. Our goal is to understand how erythrocytes geometrical changes along the capillary can affect its capacity to capture oxygen.

The first part of this document presents the model chosen for the red blood cells along with the numerical method used to determine and follow their shapes along the capillary. The membrane of the red blood cell is complex and has been modelled by an hyper-elastic approach coming from \cite{mills}. This camera method is then validated and confronted with a standard ALE method. Some geometrical properties of the red blood cells observed in our simulations are then studied and discussed.

The second part of this paper deals with the modeling of oxygen and hemoglobin chemistry in the geometries obtained in the first part. We have implemented a full complex hemoglobin behavior with allosteric states inspired from \cite{czer}.\\\\

Les globules rouges ou \'erythrocytes sont des cellules biconcaves et sont form\'ees d'une membrane contenant le cytosol, fluide ayant une grande concentration en h\'emoglobine. Cette membrane est tr\`es d\'eformable et permet \`a ces cellules de traverser des passages \'etroits comme les capillaires dont les diam\`etres sont souvent plus petits que ne l'est le diam\`etre des globules rouges. Les \'erythrocytes transportent l'oxyg\`ene dans l'organisme gr\^ace \`a la proteine d'h\'emoglobine qui dispose d'une grande affinit\'e pour l'oxyg\`ene. La capacit\'e des globules rouges \`a capturer et lib\'erer l'oxyg\`ene est ainsi un facteur d\'eterminant de leur efficacit\'e. Dans cet article, nous nous restreindrons \`a l'\'etude d'un capillaire pulmonaire o\`u les globules rouges capturent l'oxyg\`ene. Nous d\'eveloppons une m\'ethode de cam\'era de fa\c{c}on \`a \'etudier num\'eriquement le comportement de globules rouges le long d'un capillaire. Notre but est de mieux comprendre comment les \'eythrocytes se d\'eforment au long du capillaire et comment cela affecte leur capacit\'e \`a capturer l'oxyg\`ene.  

La premi\`ere partie de ce document pr\'esente le mod\`ele que nous avons choisi pour repr\'esenter le globule rouge, ainsi que la m\'ethode num\'erique utilis\'ee pour d\'eterminer et suivre la forme du globule le long du capillaire. La membrane du globule rouge est complexe et a \'et\'e mod\'elis\'ee par une approche hyper-\'elastique issue de \cite{mills}. Cette m\'ethode de cam\'era est ensuite valid\'ee et confront\'ee \`a une m\'ethode ALE standard. Enfin, quelques propri\'et\'es g\'eom\'etriques du globule rouge r\'esultant de nos simulations sont \'etudi\'ees.

La deuxi\`eme partie de ce papier traite de la mod\'elisation de la chimie entre l'h\'emoglobine et l'oxyg\`ene dans des g\'eom\'etries issues de la premi\`ere partie. Un mod\`ele complet du comportement de l'h\'emoglobine a \'et\'e impl\'ement\'e, en particulier faisant intervenir deux \'etats allost\'erique des mol\'ecules d'h\'emoglobine. Ce mod\`ele s'inspire de \cite{czer}.
\end{abstract}

\section*{Introduction}

Human respiration and more generally mammal respiration, is a complicated efficient system that uses oxygen amongst other products in order to provide energy to the body. One of the function of the respiration is thus to bring oxygen from air into the mitochondria where the major energy source of the body, the ATP, is synthesized. Oxygen travels a complex exchange system in different media. Ambient air is first brought on an air/blood exchange surface (the aveoli) through the lung, where oxygen is extracted from air. The fractal like structure of lungs (dichotomous tree of twenty-three generations) permits the exchange surface to be large enough ($\sim 100 \; m^2$) to fit energy needs of the body. The particular structure and functioning of the lung are very important in respiration and have been intensively studied, see for example \cite{felici,mauroy} but mostly independently of blood behavior. The next step of respiration for oxygen is to cross the exchange surface, to reach the red blood cells in the capillaries (smaller blood vessels) and to react with hemoglobin in order to be carried away from the lung towards the cells and mitochondria through an other branched complex structure, the vascular network, which also has been studied intensively, see for example \cite{quar}.

In this paper, we are interested in the modeling of the step of oxygen transport that takes place in the pulmonary capillaries. There, oxygen crosses the membrane between the alveola and the capillary and dissolves into plasma, where oxygen is transported by a convection/diffusion process towards the red blood cells. Next, oxygen crosses the membrane of the red blood cell and diffuses inside it (in the cytosol), then meeting an hemoglobin molecule, a chemical reaction can occur and oxygen is trapped by the hemoglobin \cite{weibel}. 

Such a modeling can lead to a better understanding of the phenomena involved in oxygen exchanges at capillary level. In particular, some pathologies can affect these exchanges by modifying critical parameters. For example, the rigidity of the red blood cells membrane can be modified (paludism, diabetes) and lead to less efficient shapes in terms of capacity to cross capillaries or capacity to catch oxygen. Similarly, the chemical properties of hemoglobin are very sensitive to blood parameters (as pH) and global consequences on oxygen transport in blood could be determined. 

The convection/diffusion process involved in oxygen transport in the capillaries is highly dependant on the shape of the structures (capillary and red blood cells) but also on the velocity field of the plasma. Hence the determination of these are required in order to correctly model oxygen displacements. The capillaries diameters are often much smaller than the red blood cells' ones, but thanks to their highly deformable membrane the red blood cells shapes can change under external constraints and they can cross the smaller vessels (parachute shape).

In order to deal with these different points, a numerical camera method is used to follow the red blood cells deformation along a straight capillary. This method is based on the standard ALE method for fluid structure interaction but is written in a moving frame. The benefits of this approach is that it is not limited by mesh skewing and the whole capillary can be covered. The mechanics of the membrane of the red blood cells is modeled by a Yeoh hyper-elastic model from \cite{mills} and validated. Some properties of red blood cells (velocities, shapes in the capillaries) are studied along with apparent hydrodynamical resistances of capillaries crossed by red blood cells.

The second step of this paper is to add to the previous modeling the behavior of oxygen in the plasma and in the cytosol, and the behavior of hemoglobin in the cytosol. Two physical processes are involved: displacement (convection and diffusion for oxygen, diffusion only for hemoglobin) and chemistry (oxygen and hemoglobin reactions). A realistic model of hemoglobin chemistry is used \cite{czer} and implemented numerically. Some preliminary results are presented, mostly on the consequences of capillaries diameter on the red blood cells capacity to capture oxygen.

Note that all simulations have been performed using Comsol Multiphysics 3.3a and that most of our calculations are 2D-axisymmetric.

\section{Fluid structure interaction: modeling}

\subsection{Geometries}

\subsubsection{Red blood cells}

\begin{figure}[h]
\centering 
\includegraphics[height=2.5cm]{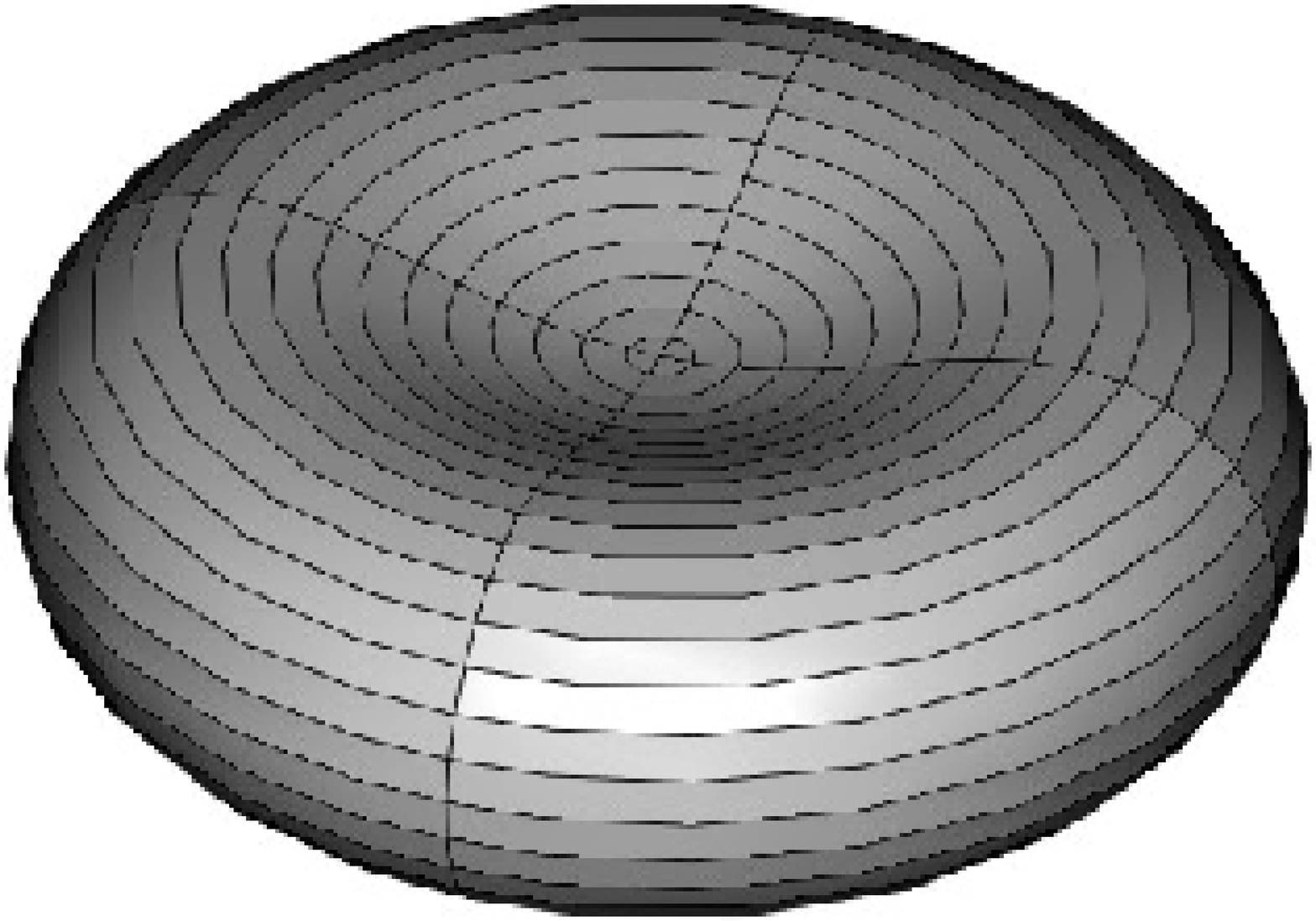}
\qquad
\includegraphics[height=2.5cm]{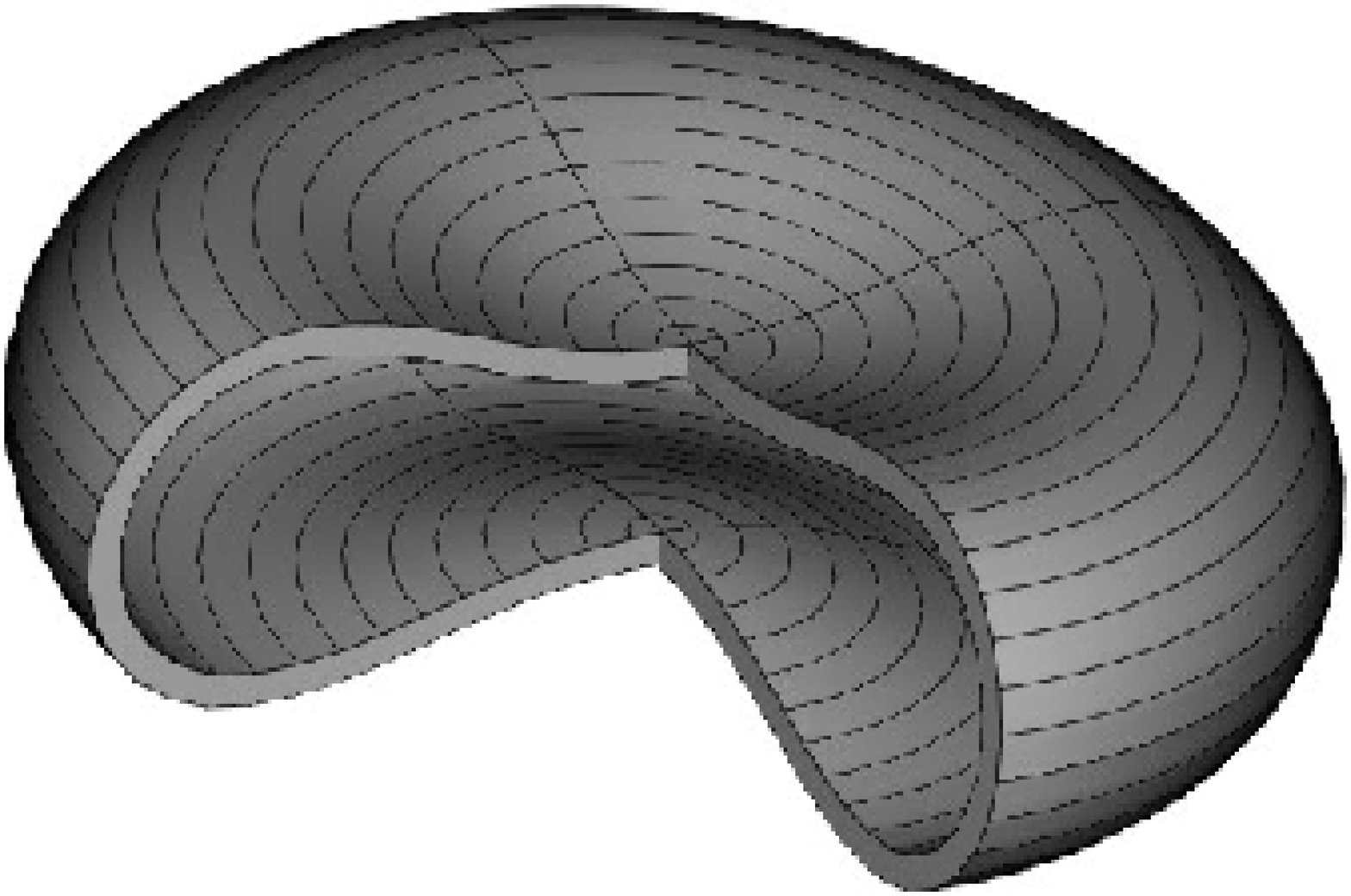}
\caption{Geometrical model used for the red blood cells. The radius of the red blood cell is $7.34 \; \mu m$, its thickness ranges from $1.4$ to $2.4 \; \mu m$ (data from Weibel \cite{weibel}). The membrane thickness is $200 \; nm$.}
\label{RBC}
\end{figure}

The red blood cell can be considered as vesicle with a membrane enclosing a fluid, the cytosol. The shape of the red blood cell is particular and known to minimize the bending energy of the membrane \cite{canham} amongst other properties \cite{uzoi}, see figure \ref{RBC}. In the following work, the geometrical parameters of the red blood cells have been chosen accordingly to Weibel's data \cite{weibel} for human erythrocytes.

The cytosol can be considered as a viscous fluid and contains a high concentration of hemoglobin. It will be considered as a Newtonian viscous fluid: its density corresponds to that of water while its viscosity is eight times water viscosity \cite{glenor}.

The membrane of the red blood cells is a superposition of a bilipidic layer and of an actin mesh. Its thickness is of the order of one or two hundreds of nanometers \cite{weibel}. 

The mechanical behavior of the membrane is complex and not yet well understood. However some interesting models have been developed \cite{arslan,mills,pnas}. More precisely, in \cite{mills}, a relatively simple hyper-elastic model for the membrane has been applied successfully to recreate numerically optical tweezers experiments. This hyper-elastic model has been developed by Yeoh in \cite{yeoh} and uses the following strain energy:

\begin{equation}
W = \frac{G_0}{2} (\lambda_1^2+\lambda_2^2+\lambda_3^2-3) + C (\lambda_1^2+\lambda_2^2+\lambda_3^2-3)^3
\end{equation}

where $G_0$ is the initial shear modulus of the membrane and $C$ a constant estimated to $G_0/30$ in order to fit optical tweezers experiments, see \cite{mills}. Note that in the literature, data are often given in term of in-plane shear modulus: $\mu_0 = G_0 h_0$ where $h_0$ is the initial thickness of the membrane. In the following, $\mu_0$ has been chosen to $11.3 \; \mu N / m$, see \ref{verif}.

\subsubsection{Reproducing Mills et al experiments}
\label{verif}

In order to validate the membrane model, a numerical reproduction of Mills et al experiments \cite{mills} has been performed. These experiments consist in attaching a metallic bead at each extremity of the red blood cell and to use magnetic sinks in order to control the positions of the beads (optical tweezers). This method gives access to the force applied to the beads and thus it is possible to deduce some mechanical properties of the membrane if measuring the deformations against the force applied.

\begin{figure}[h]
\centering 
\includegraphics[height=2.5cm]{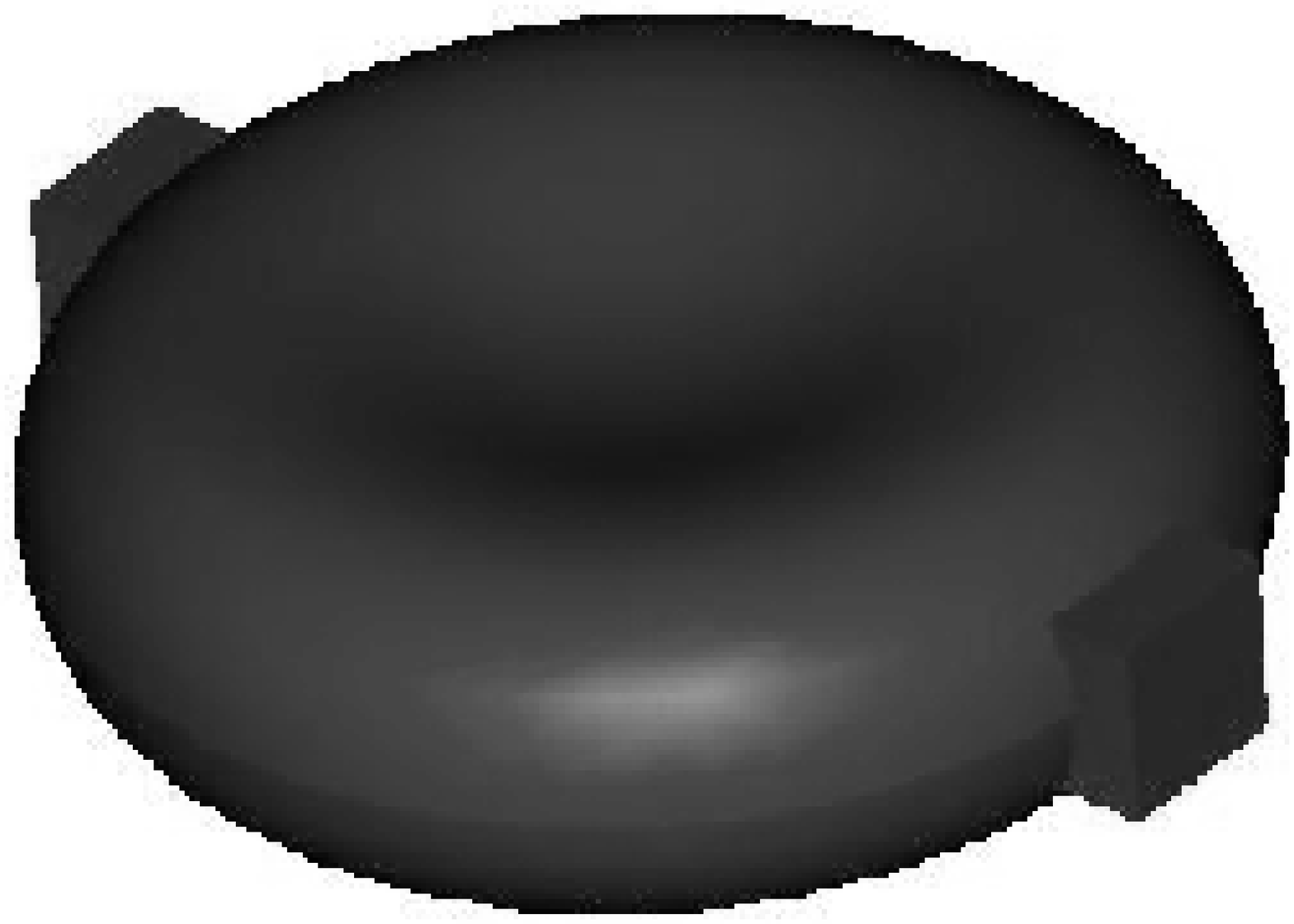}
\qquad
\includegraphics[height=2.5cm]{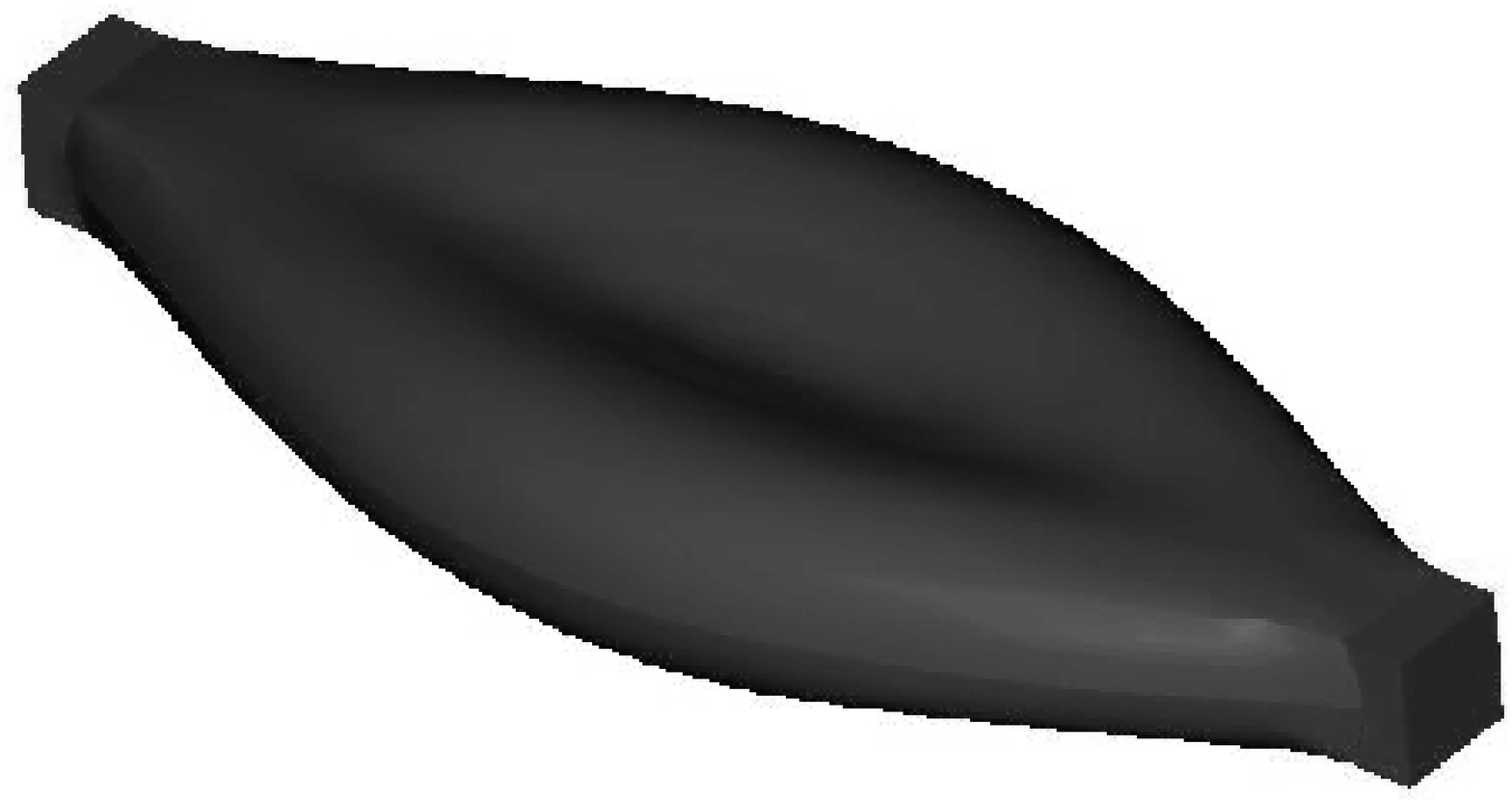}
\caption{Left: initial globule geometry, with ``beads'' at each side. Right: deformed globule with a force of $120 pN$.}
\label{geommills}
\end{figure}

In order to build a numerical equivalent, we use our three dimensional red blood cell. Two cubic rigid ``beads'' have been attached on its membrane. One of these beads will be fixed while the other will be submitted to a force. The inside of the red blood cell is considered as a Stokes fluid with previously given characteristics. Note that because the fluid is assumed incompressible, the red blood cell keeps a constant volume.
We use an ALE fluid-structure interaction model. Figure \ref{geommills} shows the initial geometry and a typical deformed globule under stress.   

\begin{figure}[h]
\centering 
\includegraphics[height=5cm]{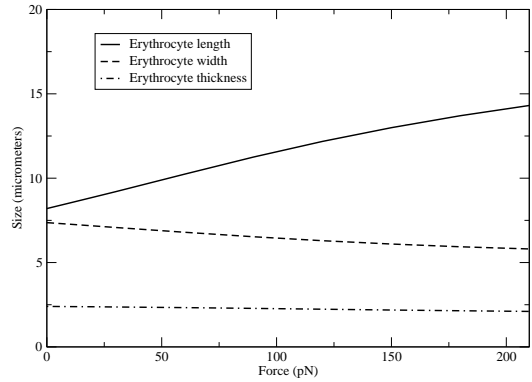}
\caption{Deformation of the red blood under optical tweezers stress, numerical results.}
\label{millsres}
\end{figure}

We have measured lengths, widths and thicknesses of the red blood cell for a range of forces going from $0$ to $210 pN$ with steps of $30 pN$. We have found a better collusion between our results and Mills et al results for an in-plane shear modulus of $11.3 \mu N / m$ and with the parameter $C$ equal to $G_0/30$. The corresponding results of our numerical simulations are shown on figure \ref{millsres}.
In the following, we will keep these parameters constant to these validated values.

\subsubsection{Capillary}

\begin{figure}[h]
\centering 
\includegraphics[height=3cm]{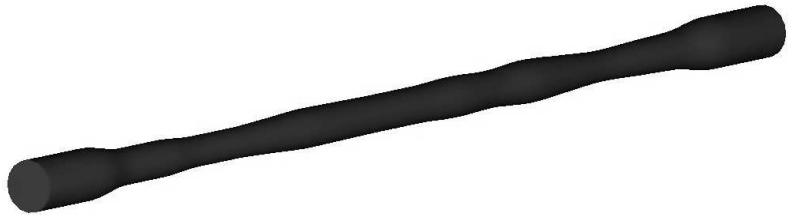}
\caption{Example of a capillary}
\label{capill}
\end{figure}

We will assume at first approximation that a capillary is topologically equivalent to a cylinder whose axis can be considered as a straight line. In the camera method developed below, this will simplify the movement of the camera, limiting it to a translation along the capillary axis. The walls of the capillary will be assumed unmoving and rigid. The shape of the capillary will be chosen according to measures done in \cite{di1}. The typical length of such a vessel is of the order of one millimeter and its radius can range from $5 \mu m$ to $15 \mu m$. Figure \ref{capill} shows a geometrical model of capillary that will be used in the next sections.

The plasma will be considered as a viscous Newtonian fluid with same density than water, but with twice its viscosity. Because capillaries are far from the heart, they do not feel heart beating and thus we can consider that the plasma circulates with a constant flow rate along the capillary. Moreover the velocities are of the order of the millimeter per second and the Reynolds number $\rho v d / \mu$ can be estimated to $1000 \times 10^{-3} \times 10^{-5} / 2.10^{-3} = 0.005$. Thus Stokes equations are a good approximation of plasma behavior in the capillary.  

\subsection{Equations}

Because red blood cells are travelling through the whole capillary, the coordinates map induced by their displacement in a standard ALE method will be highly skewed. Direct numerical simulations using finite elements will thus be very difficult because of highly deformed elements. This could be bypassed by regular remeshings but this method is relatively costly and relatively difficult to set up.

Hence, considering the typical shapes of the studied capillary (that is assumed to be topologically equivalent to a cylinder and to keep its axis along a straight line) and considering the fact that our objects of interest are the red blood cells, a camera method seems advantageous. This method consists in focusing only on the portion of the capillary where the red blood cells are and to displace focus together with red blood cells.  

Numerically speaking, this consists in meshing a small portion of the capillary, in making the mesh move along it at red blood cells mean velocity and also in making the mesh fit capillary local shape. Note that because we do not need to mesh the complete capillary, this method will also benefit of a reduced number of elements.

The drawbacks of this method are located on boundaries: 

\begin{itemize}
\item the coordinates map induced by red blood cells deformations must also include geometrical changes of the capillary walls along its length
\item the boundary conditions for the plasma on the non-walled boundaries are not obvious and need to be studied more closely
\end{itemize}

%We will now write the fluid structure interaction equations using the camera method. 

Note that all bearings involved in the following will be built thanks to the Winslow smoothing method \cite{winslow}, which robustness against mesh elements inversion is better than Laplace method.

\subsubsection{Standard method}

First we will write the standard method, namely the standard fluid structure equations in ALE formulation on the full capillary. 
%Because it is topologically equivalent to a cylinder, we will assume that our reference domain $\Omega$ is a cylinder. 
The capillary domain will be called $\Omega$ and its straight axis will correspond to one of the reference frame axis. In the following, this particular direction will be recognized amongst the other by the subscript $c$. The reference coordinate system will be denoted $(x)$, and the basis vectors will be noted $(e_i)_i$. Thus the cylinder axis is directed by $e_c$ and the position on this axis referenced by $x_c$.

\begin{figure}[h]
\centering 
\includegraphics[height=3.5cm]{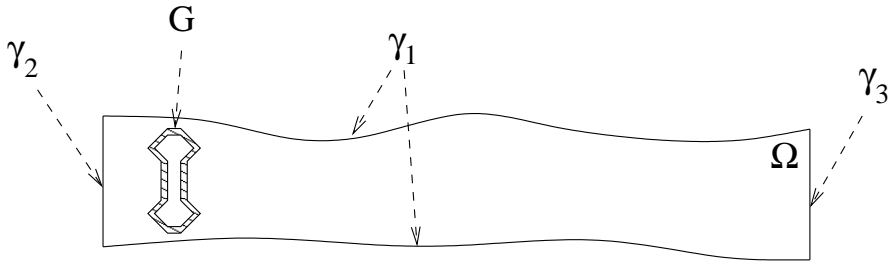}
\caption{Reference geometry for standard ALE method.}
\label{ALEcap}
\end{figure}

Let $G \subset \Omega$ be the membrane of the red blood cells and $F = \Omega \backslash G$ the reference domain for the fluid. Note that $F$ corresponds not only to the outside of the red blood cells (plasma), but also to their insides (the cytosol), these two fluids have the same density (similar to water) but a different viscosity. Let $\gamma_i$ ($i=1,2,3$) be the capillary walls as shown on figure (\ref{ALEcap}) such that $\partial \Omega = \gamma_1 \cup \gamma_2 \cup \gamma_3$.

The standard fluid structure equations in ALE formulation are:

\begin{equation}
\begin{array}{ll}
\rho_g \frac{\partial^2 u}{\partial t^2} - div(\sigma(u)) = 0 & \text{ on } G\\
&
\\
\left\{
\begin{array}{l}
\rho_f \frac{\partial v}{\partial t} - \mu_f \bigtriangleup v + \nabla p = 0 \\
div(v) = 0
\end{array}
\right.
&
\text{ on } F_t = \psi(F,t)
\end{array}
\label{ale}
\end{equation}

with boundary conditions:

\begin{equation}
\begin{array}{l}
\begin{array}{ll}
\sigma(u).n = ``\psi^{-1}(\mu_f \nabla v.n - p.n)'' & \text{ on } \partial G\\
&\\
\end{array}
\\
\left\{
\begin{array}{ll}
v = \frac{\partial u}{\partial t} & \text{ on } \partial G_t = \psi(\partial G,t)\\
v = 0 & \text{ on } \psi(\gamma_1,t) \; =\gamma_1\\
v = v_e & \text{ on } \psi(\gamma_2,t) \; =\gamma_2\\
\mu_f \nabla v.n - p n = 0 & \text{ on } \psi(\gamma_3,t) \; =\gamma_3
\end{array}
\right.
\end{array}
\end{equation}

and $\psi = Id + \varphi$, where $\varphi$ is a bearing of $u$ on whole $\Omega$ with boundary constraints $\varphi = u(.,t)$ on $\partial G$ and $\varphi.n = 0$ on $\partial \Omega$. The constraints equality is done on the reference frame and the notation $''\psi^{-1}(...)''$ means that the constraints tensor of the fluid defined in the deformed frame is pulled back in the reference frame. The boundary condition on $\partial G$ is the displacement due to the deformation of the membranes of the red blood cells, this deformation comes from equations (\ref{ale}). The boundary condition on $\partial \Omega$ states that the capillary walls are unmovable but allows the coordinates of the deformed frame to slide along the walls of the capillary. The coordinate in the deformed frame $\psi(\Omega,t)$ will be noted $y$ and $y=\psi(x,t)$.

\subsubsection{the camera method}

We will now restrict the domain for the camera method: we will now consider a coordinates system $(X)$ directed by $(E_i)_i$ and assume our reference frame is a cylinder $C$ which axis is along $X_c$. Its length along $X_c$ is $L>0$ ($L$ is smaller than capillary length). This cylinder will fit the local capillary geometry while it follows the red blood cells along the capillary. The deformed cylinder at time $t$ will be called {\it the camera} and will be noted $C_t$. The domain defined by the red blood cells membranes is called $G \subset C$. 

%According to the ALE method, the membrane deformation will be calculated in the reference frame $C$ and the fluid in the deformed frame $C_t$.

The camera will follow the red blood cells in such a way their mean position is always at the center of the view of the camera. Hence the position along the capillary of the camera center can be calculated: 

\begin{equation}
m_c(t)=m_c^0+\frac{\int_G u_c(.,t)}{\int_G 1}
\end{equation}

The displacements $u$ are still calculated in the reference frame on the membrane domain $G \subset C$ and will not differ from the standard ALE method. In order to rebuild the geometry of the capillary containing the deformed globules, it is necessary to first integrate the deformed globules into the non-deformed capillary and then to make the cylinder wall fit the capillary without changing the shapes of the globules. This will be done using two successive bearings $Y=A(X,t) = X + \alpha(X,t)$ and $Z = B(Y,t) = Y + \beta(Y,t)$:

\begin{equation}
\begin{array}{lll}
\left\{ 
\begin{array}{ll}
\alpha = 0 & \text{ on } \Gamma^i \; (\forall i) \\
\alpha = u(.,t) - m_c(t).E_c & \text{ on } \partial G \\
\end{array}
\right.
&
\text{ and }
&
\left\{
\begin{array}{ll}
\beta.n(Y) = g(Y+\beta(Y,t)+m_c(t).E_c) & \text{ on } A(\Gamma^1,t) \\
\beta = 0 & \text{ on } A(\partial G,t) \text{ and } A(\Gamma^i,t) \; (\forall i\neq 1) \\
\end{array}
\right.
\end{array}
\end{equation}

\begin{figure}[h]
\centering 
\includegraphics[height=3.5cm]{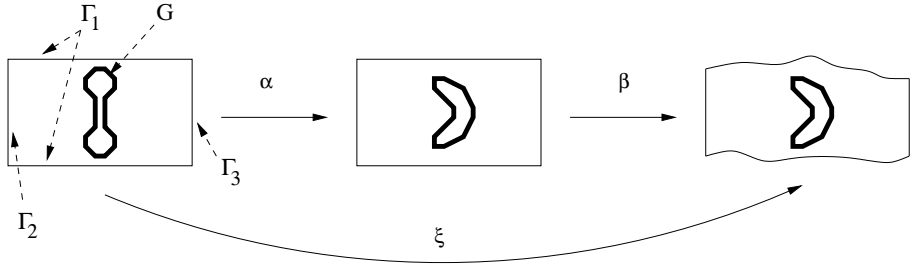}
\caption{Domains notations and bearings involved in the camera method.}
\end{figure}

Thanks to the definition of $m_c(t)$, the transformation $A$ preserves the mean position of the red blood cells to $0$ on the $X_c$ axis.

Stokes equations (velocity will be called $w$ and pressure $\tilde{p}$) will be calculated in $(Z)$ coordinates in order to be solved in the correct geometry. The transformation $A$ corresponds to the movement of the camera and will thus induce an additive transport term in the Stokes equations: $- (\frac{d m_c}{dt}.\nabla)v$. However, because of the small Reynolds numbers involved and thanks to the fact that $\frac{d m_c(t)}{dt} \sim v$, this supplementary term is negligible the same way $(v.\nabla) v$ is. Moreover, the transformation $B$ does not modify Stokes equations.

Finally, the succession of these two bearings can be reduced to a single bearing $\Xi(X) = X+\xi(X,t) = X + \alpha(X,t) + \beta(X+\alpha(X,t),t)$. The boundary conditions for $\xi$ are:

\begin{equation}
\left\{
\begin{array}{ll}
\xi = u(.,t) - m_c(t).E_c & \text{ on } \partial G_t\\
\xi.n(X) = g(X+\xi(X,t)+m_c(t).E_c) & \text{ on } \Gamma^1\\
\xi.n = 0 & \text{ on } \Gamma^2 \text{ and } \Gamma^3
\end{array}
\right.
\end{equation}

The boundary conditions for fluid equations will be the same as for the standard ALE method concerning the capillary wall and the globules wall (non slip conditions). However, they should be studied for what concerns the inlet ($\Xi(\Gamma_2)$) and the outlet ($\Xi(\Gamma_3)$) of the camera domain. In the standard ALE method, a Dirichlet condition on velocity was imposed at the outlet $\gamma_3$ and a zero pressure condition was introduced at the inlet $\gamma_2$. The Dirichlet condition can be easily used at camera outlet considering the very low Reynolds numbers: a Poiseuille profile should be a good approximation of the velocity at camera outlet, all the most if the border is not too close to the red blood cells. At inlet ($Z_c = -L/2$) however, the correct pressure condition should be $\tilde{p}(Z,t) = p(y,t)$ with $Z$ such that $Z_c=-L/2$ and $y$ such that $y_c=m_c(t)-L/2$. Again, thanks to the low Reynolds numbers, we can assume that pressure is constant on the section $y_c = m_c(t)-L/2$ and impose $\tilde{p}$ to fit the mean value on this section namely $p_0(t)$. However this value is only time dependant and because Stokes equations only depends on $\nabla \tilde{p}$, making the variable change $q(Z,t) = \tilde{p}(Z,t) - p_0(t)$ will not change the equations and will allow us to impose $q=0$ on the camera inlet. 

Finally, the fluid constraints on red blood cells membranes $-p.n + \mu_f \nabla u.n$ is modified only by the relative pressure change: because this change is global it does not affect the mechanics equations.

The errors done relatively to the standard approach are due to boundary conditions simplifications and to the suppression of the new transport term. We will do numerical comparisons in order to show the quasi equivalence between both methods and show that, in the contrary of the standard approach, the camera method can follow the red blood cells all along the whole capillary.

\section{Fluid structure interaction: numerical results}

\subsection{Comparison between the standard method and the camera method (2D axisymmetric)}

\begin{figure}[h]
\centering 
\qquad
\includegraphics[height=1.5cm]{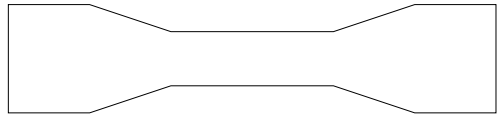}
\caption{Capillary geometry used in this section, the red blood cell enters to the left part where radius is $4.5 \mu m$, the full structure is $1.5 mm$ long and the constricted part is $0.5 mm$ long with a radius of $3 \mu m$.}
\label{compgeom}
\end{figure}

In order to validate the camera method, a comparison has been made between the standard ALE method and its pendant in camera method. In order to limit the calculation time, a 2D axisymmetric case has been used. In this example, a red blood cell has been put inside a plasma flow through an axisymmetric bottleneck. At first the radius of the capillary is larger ($4.5 \mu m$) than that of the globule ($3.6 \mu m$) but it is smaller in the constriction ($3 \mu m$), see figure \ref{compgeom}.\\

\begin{figure}[h]
\centering 
\includegraphics[height=5cm]{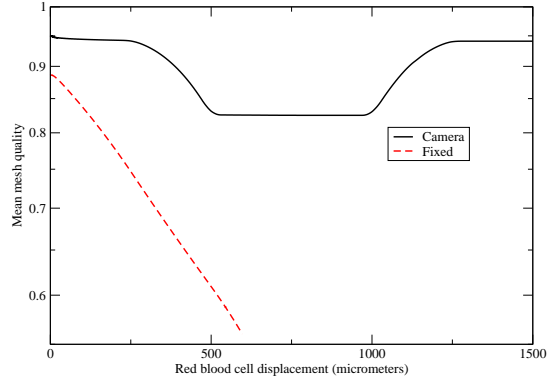}
\caption{Comparison between standard ALE method (red dashed curve) and camera ALE method (black continuous curve) in a bottleneck like ``capillary'': mesh qualities.}
\label{compmesh}
\end{figure}

Because of its design, the camera method induces a far smaller distortion of the mesh and can follow easily the red blood cell along the whole geometry. The mesh quality for the camera method is only dependant on the radius of the capillary and not on the position of the red blood cell. On the contrary, mesh quality is decreasing linearly in the standard method, and the numerical simulation stops before the end of the bottleneck (at time $1.74 s$ about $600 \; \mu m$ from entry) because of highly skewed elements. Mean mesh qualities versus globule positions are represented on figure \ref{compmesh}.\\\\

Some parameters have been checked in order to compare the differences between both methods. We have measured the displacement of the red blood cell and also its aspect ratio along time in both simulations. The data show a very good agreement between both methods on the range of convergence of the standard method. There is however a slight difference for the red blood cell aspect ratio in the last times of convergence of the standard method that is a consequence of some of its elements being highly skewed: this lead to a bad convergence of the problem solution.\\   

\begin{figure}[h]
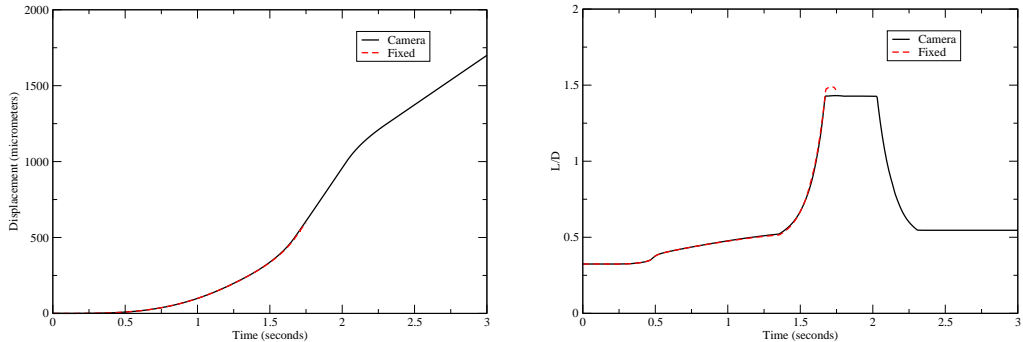

\centering
\includegraphics[height=4.5cm]{depm2.eps}
\qquad
\includegraphics[height=4.5cm]{LoD2.eps}
\caption{Comparison between standard ALE method (red dashed curve) and camera ALE method (black continuous curve) in a bottleneck like ``capillary''. Left: position of the red blood cell in the capillary. Right: length over diameter ratio of red blood cell.}
\label{comp}
\end{figure}

\subsection{Going along a capillary}

Being able to follow red blood cells along a capillary gives access to a large quantity of data that needs to be studied in lots of different contexts. We will present here some of these interesting results but keeping in the wide context: ``how do it feel being a red blood cell in a capillary ?''.

In this section, and as before, we will limit us to a 2D axisymmetric problem in order to limit calculations times.

\begin{figure}[h!]
\centering 
\includegraphics[height=5cm]{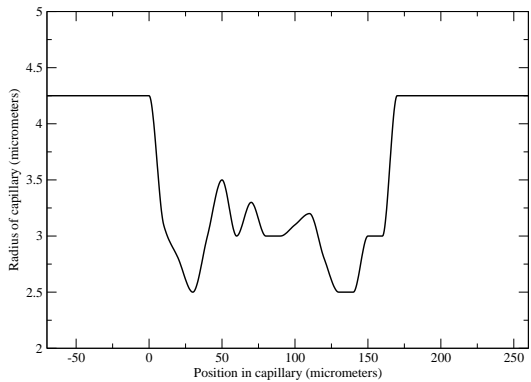}
\caption{Capillary radius.}
\label{hcap}
\end{figure}

We will do our study on an infinite capillary along axis $x_c$ (measures will be done in meters). The vessel diameter is equal to $8.5 \mu m$ on $x_c \in \; ]-\infty,0[ \; \cup \; ]170.10^{-6},+\infty[.$ and is variable on the range $x_c \in \; [0,170.10^{-6}]$. The variations chosen are close to the one observed on an in-vivo rat capillary in \cite{jeong} and are shown on figure \ref{hcap}. We will call $C_v$ the portion of capillary corresponding to the region of diameters variation, namely with $x_c \in \; [0,170.10^{-6}]$.

Depending on the case, we will put one, three or five red blood cells in the capillary. The initial distance between them has been put to $10 \mu m$ in order to avoid collisions. This value is higher than the observations that gives a number closer to $5 \mu m$ (data calculated from a red blood cells volume fraction of $0.4$ in blood, see \cite{quemada,weibel}). So that they have a stabilized shape at the entry of the portion of capillary $C_v$, the initial mean position of the globules has been put to $x_c=-650 \mu m$. Plasma flow through a capillary section of $8.5 \mu m$ diameter has been chosen to correspond to a mean velocity of $0.5 mm/s$, in the simulations the flow is maintained constant whatever the position of the camera, in particular it implies velocity increase when capillary radius decreases.

\subsubsection{What shape?}

\begin{figure}[h!]
\centering 
\includegraphics[height=2cm]{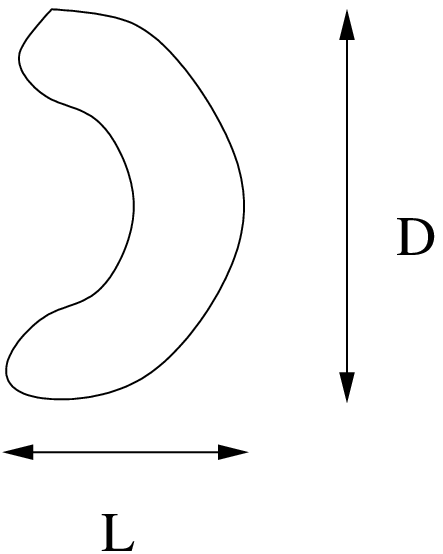}
\qquad
\includegraphics[height=5cm]{LoD.eps}
\caption{Length (L) over diameter (D) ratio of the middle red blood cell.}
\label{LoD}
\end{figure}

Under plasma flow stress, the red blood cells are moving forward the capillary, and because of the non homogeneous shape of the flow on a capillary section, the globules shape changes to the typical parachute shape, pointing in the direction of the flow. Moreover the globules go through skewer shapes when the radius of the capillary decreases. In order to estimate the distortion of red blood cells, we will use the aspect ratio as in \cite{jeong}. It is defined by the ratio of cells length along the capillary axis over their diameter, see figure \ref{LoD}. Under no stress, this aspect ratio is $2.4/7.34 \sim 0.33$.\\\\

\begin{figure}[h!]
\centering
\begin{minipage}[c]{0.40\linewidth}
\centering
\includegraphics[height=2.5cm]{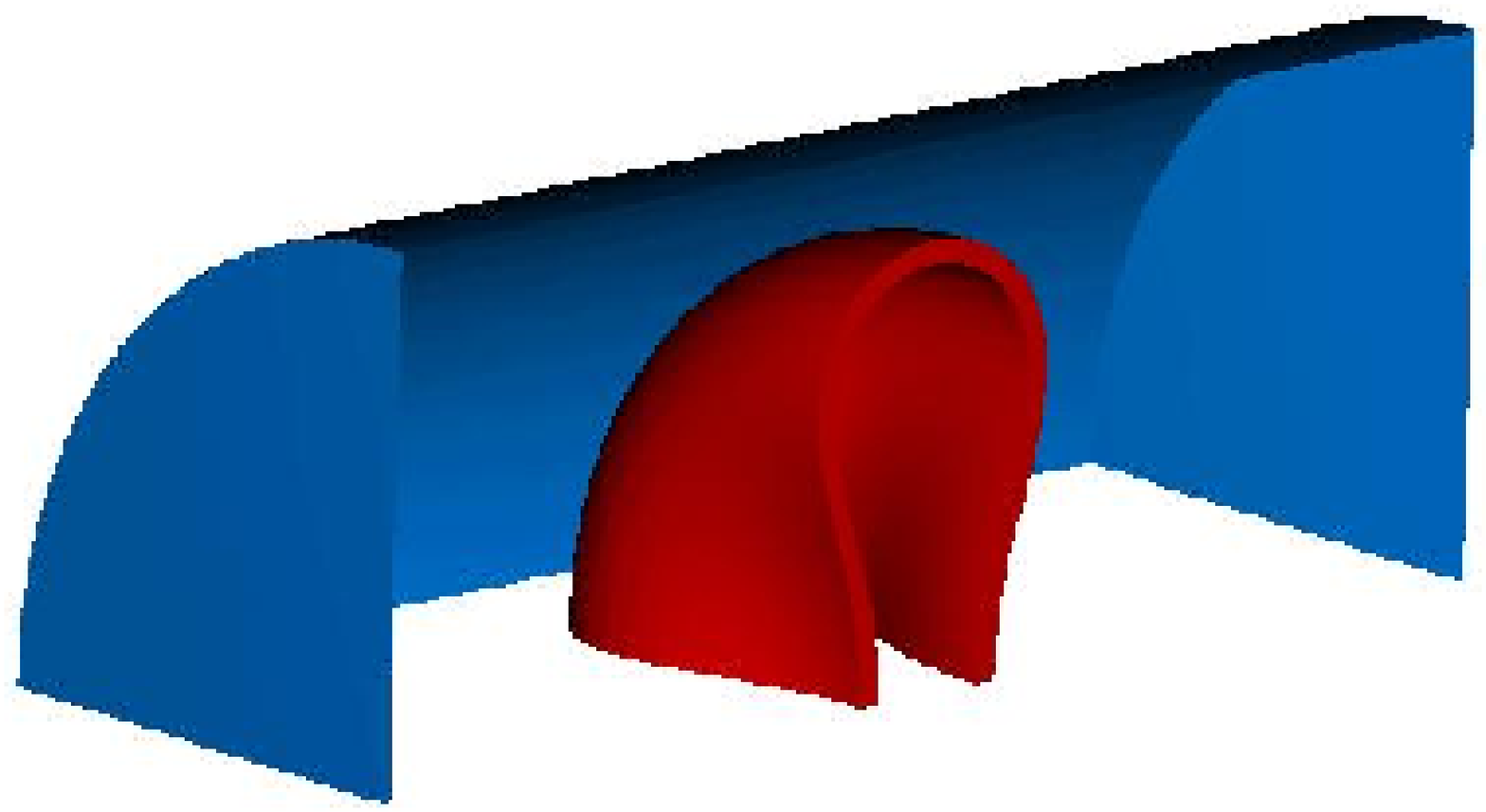}\\
{\tiny (a) capillary diameter of $8.5 \mu m$, no plasma stress. Aspect ratio of $0.33$.}
\end{minipage}
\begin{minipage}[c]{0.40\linewidth}
\centering
\includegraphics[height=2.5cm]{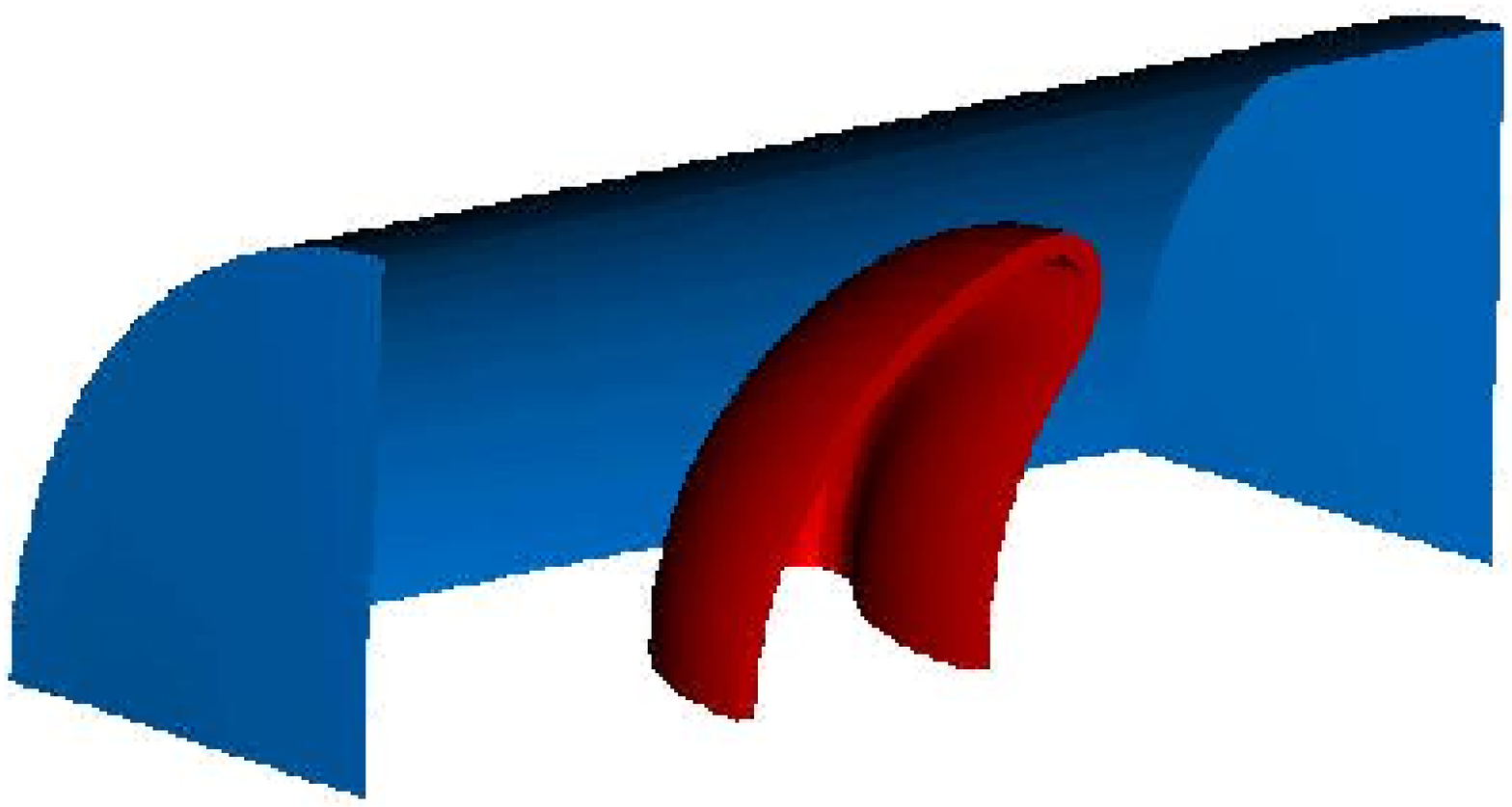}\\
{\tiny (b) capillary diameter of $8.5 \mu m$. Aspect ratio of $0.55$.}
\end{minipage}

\begin{minipage}[c]{0.40\linewidth}
\centering
\includegraphics[height=2.5cm]{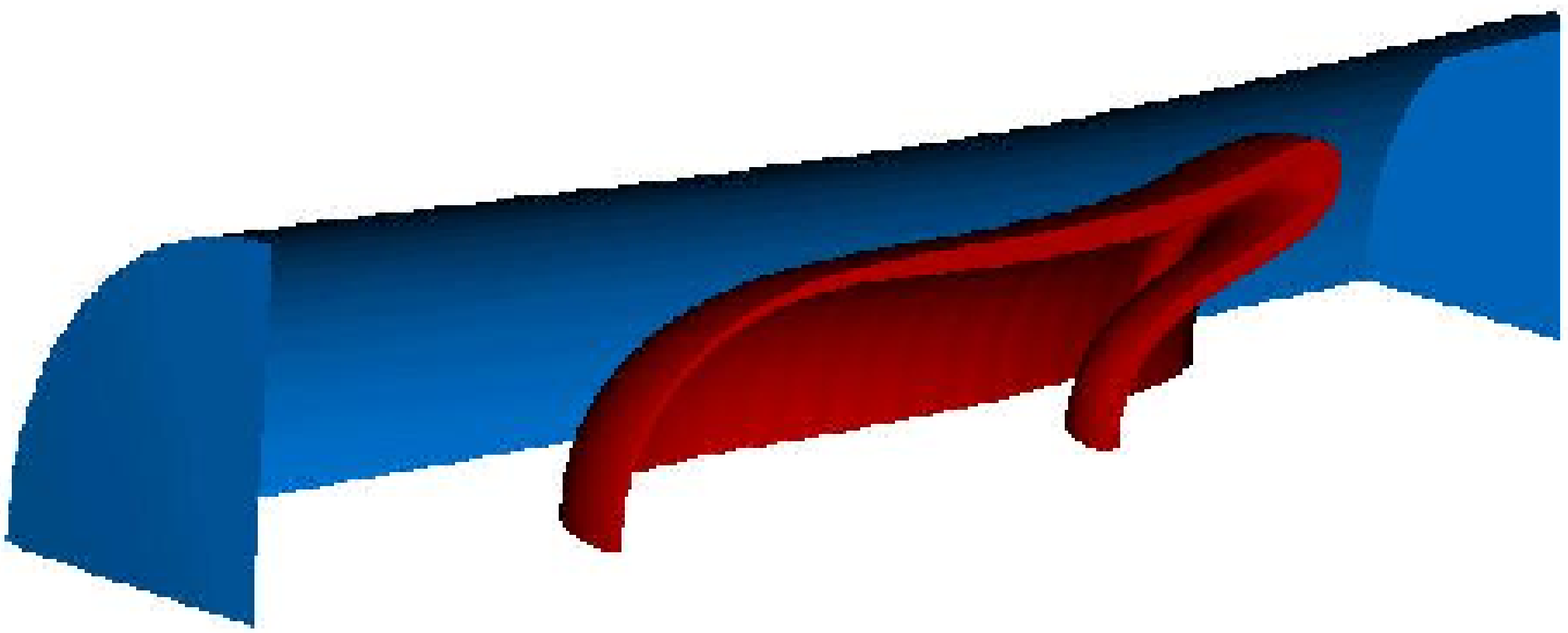}\\
{\tiny (c) capillary diameter of $5.5 \mu m$. Aspect ratio of $1.75$.}
\end{minipage}
\begin{minipage}[c]{0.40\linewidth}
\centering
\includegraphics[height=2.5cm]{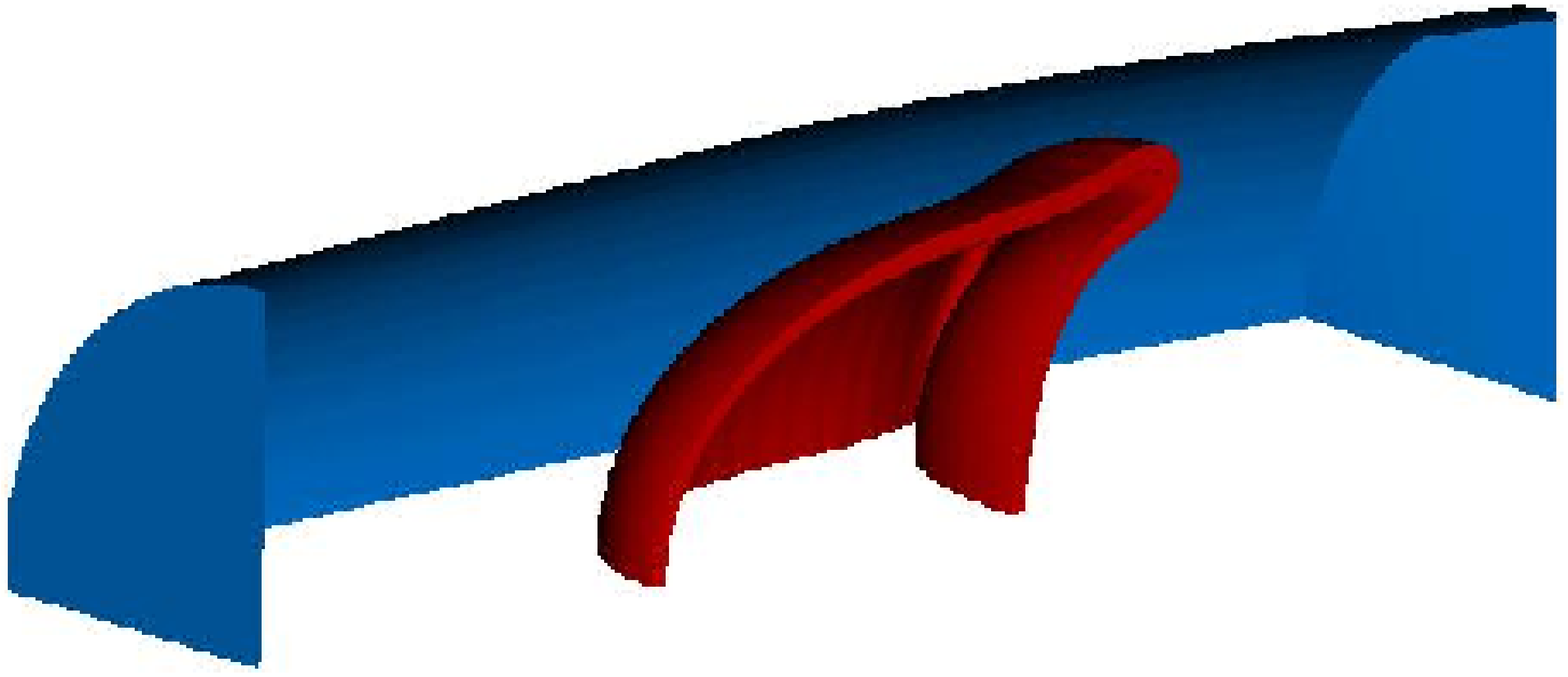}\\
{\tiny (d) capillary diameter of $6.5 \mu m$. Aspect ratio of $1$.}
\end{minipage}

\caption{(a) Initial shape of the red blood cell. (b),(c),(d) Shapes of red blood cell under flow stress for different values of capillary radius.}
\label{globshapes}
\end{figure}

Figure \ref{LoD} shows the different curves of aspect ratio depending on the number of red blood cells involved in the calculation. Red blood cells aspect ratio follows radius changes of capillary. This illustrate how able to deform the red blood cells are: starting from an aspect ratio of $0.33$, they can reach almost $2$ in the narrowest parts of the capillary. These numbers are fully coherent with the measures made in \cite{jeong}, where the authors have observed similarly the red blood cells aspect ratios and have more particularly studied these changes against capillary radius variation. On figure \ref{globshapes}, typical shapes observed in our simulations have been drawn, in particular we can observe the classical parachute shape.

The presence of more than one red blood cells plays a small but visible role in aspect ratio determination for our choice of parameters (distance of $10 \mu m$ between each cell). This indicates that closer globules, as in real blood, should influence themselves, all the more if there are collisions between them.

\subsubsection{Which velocity?}

\begin{figure}[h!]
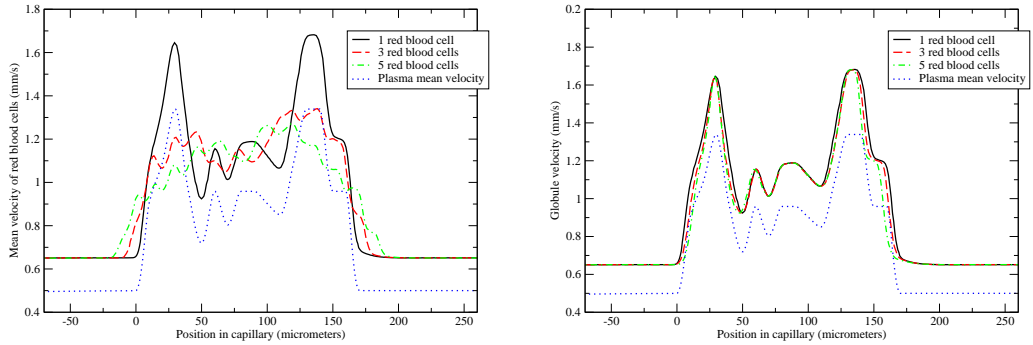

\centering 
\includegraphics[height=4.5cm]{vitm.eps}
\qquad
\includegraphics[height=4.5cm]{vitm_glob_milieu.eps}
\caption{Left: mean velocities of the red blood cells. Right: velocity of the middle red blood cells.}
\label{vitm}
\end{figure}

Because the skewed red blood cell does not fill the whole section of the capillary, it is submitted only to the larger part of plasma velocities which stands mostly in the middle of the capillary (Poiseuille profile). Thus, the mean velocity of the cells is most of the time larger to the mean velocity of the plasma. This characteristic of blood is well known \cite{eirich}. Similarly, it is interesting to observe that the gap between the cells and the capillary wall decreases with capillary radius and can indeed modify this phenomena. The velocities of plasma and the velocities of cells in different cases (one, three or five cells in the capillary) are shown on figure \ref{vitm}. Note that, for the left figure, increasing the number of cells in the capillary will induce an homogenisation of the mean velocity because each cell sees a different radius than the others. On the right part of the figure the velocities of the middle red blood cell is shown, the effects of the presence of other cells is small but exists. Hence, decreasing the initial distance between cells should induce a higher interaction, but also create collisions which are not, for now, managed by our code.\\\\

\subsubsection{Consequences on the hydrodynamical resistance of the capillary?}

By crossing the capillary, the red blood cells interfere with vessel's hydrodynamical resistance which cannot be predicted through fluid models, even non-newtonian ones. This is due to the deformability of the red blood cells along with the similar sizes of red blood cells and capillary diameters. Thus it is necessary to study the full fluid structure interaction problem in order to obtain reasonable high scales estimation of capillaries hydrodynamical resistances. In order to apply our simulations at this problem, we have obtained preliminary results on the hydrodynamical resistance. We have measured the resistance of the portion of capillary viewed by the camera against the number of cells (zero, one, three or five). To estimate resistance we have calculated the ratio from pressure drop between the inlet and the outlet over the flow in a capillary section. 

\begin{figure}[h!]
\centering 
\includegraphics[height=5cm]{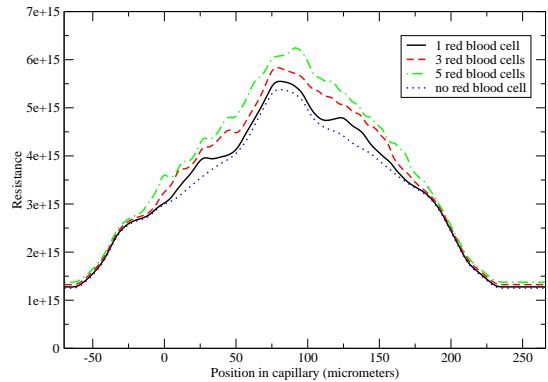}
\caption{Resistance of the portion of capillary in the camera versus camera position.}
\label{res}
\end{figure}

The results are drawn on figure \ref{res} and show that, as expected, increasing the number of cells increases the resistance. Note that at relative high diameters, the resistance is almost identical whatever the number of the globules. On the contrary, the differences increases when globules shape is changing, actually some of the fluid energy is exchanged with the membrane of the red blood cells and higher pressure drop is thus needed to ensure constant flow.\\

\section{Introducing oxygen and hemoglobin: modeling (preliminary)}

Red blood cells are designed to catch oxygen in the lung capillaries, to carry it in the whole body and to deliver it to the cells in the systemic capillaries. The way oxygen is captured and delivered is obviously critical. Actually, the time spent in the capillaries must be tuned in order to maintain the balance between oxygen delivery and oxygen needs. The chemistry of hemoglobin plays an important role in oxygen exchanges, but other actors are also involved, like blood velocity or red blood cells shapes that are determinant for good functioning of the system.

In this section we will use the $0D$ hemoglobin chemistry model developed in \cite{czer} and extend it to $2D$ axisymmetric one in the geometries obtained from the preceding sections. More precisely and because oxygen diffusion and reaction with hemoglobin can be uncoupled from red blood cells deformation, the results of the preceding sections can be considered as known data in the modeling of oxygen exchanges. 

\subsection{Modeling of hemoglobin chemistry}

\begin{figure}[h!]
\centering
\includegraphics[height=2.5cm]{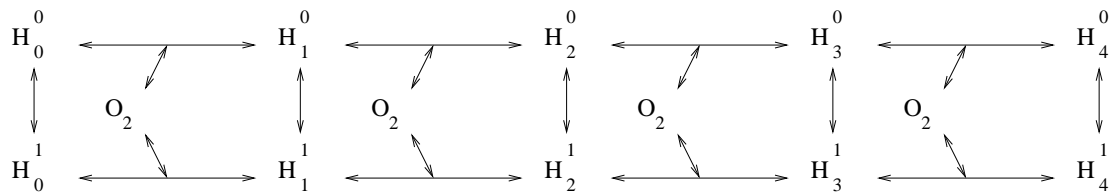}
\caption{Hemoglobin chemistry diagram (from \cite{czer}): hemoglobin has two allosteric states $T$ and $R$ (resp. exponent $0$ and $1$). Hemoglobin can go from one state to the other and can carry from 0 to 4 oxygen molecules.}
\label{O2chem}
\end{figure}

Hemoglobin is a complex molecule consisting in four hems, each of them being able to carry one molecule of oxygen. The four hems divides in two $\alpha$ and two $\beta$ hems (or chains) that have different bindings properties with oxygen. This asymmetry of the hemoglobin molecule implies the existence of two allosteric states $T$ and $R$ which will be integrated in the model.
It is not clear in which priority these chains react with oxygen, and we will follow one of the mechanisms proposed in \cite{czer} which considers that $\alpha$-chains are the first to react. With this hypothesis, the rate constants of the different chemical reactions have been determined by multiple experimental measurements which references are available in \cite{czer}. The diagram of the reactions involved in the model is shown on figure \ref{O2chem}. Note that because these experiments are not made at the temperature of human body, adjustments have been made by multiplication with a unique factor each of the constant rate of bi-molecular reactions. Once these adjustments done, we have successfully compared the resulting chemical model with known properties of hemoglobin.

In the following, we will write $h_i^j$ the concentration of hemoglobin of state $j \in \{0,1\}$ ($j=0$ corresponds to $T$ state while $j=1$ corresponds to $R$ state) carrying $i \in \{0,1,2,3,4\}$ oxygen molecules (noted $H_i^j$ on figure \ref{O2chem}).

An important number in blood studies is hemoglobin saturation, this number corresponds to the percentage of hemoglobin capacity which is used to carry oxygen. For instance, during rest regime hemoglobin saturation is $0.97$ after crossing the lungs while it is $0.75$ before. It is the ratio between the quantity of $O_2$ effectively captured over the maximum quantity of $O_2$ possible to capture. In the model chosen here, it is calculated by:

\begin{equation}
SO_2 = \frac14 \frac{(h_1^T+h_1^R)+2(h_2^T+h_2^R)+3(h_3^T+h_3^R)+4(h_4^T+h_4^R)}{(h_0^T+h_0^R)+(h_1^T+h_1^R)+(h_2^T+h_2^R)+(h_3^T+h_3^R)+(h_4^T+h_4^R)}
\end{equation}

\begin{figure}[h!]
\centering
\includegraphics[height=5cm]{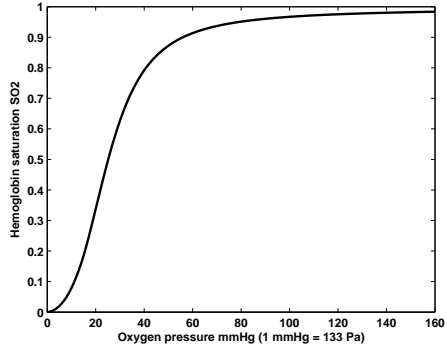}
\caption{Hemoglobin saturation given by the model versus oxygen partial pressure in the medium ($1\; mmHg = 133 \; Pa$). The curve fits well measured data given in the literature \cite{weibel}.}
\label{SO2sat}
\end{figure}

On figure \ref{SO2sat}, the saturation curve resulting from the model is shown. It has been plotted against oxygen partial pressure in the medium ($mmHg$). The steep part has an important role in oxygen exchange and its position is important, see \cite{weibel} for more details. The model fits well measurements, for example the partial pressure in oxygen needed to have a $0.5$ saturation (steep part of the curve) is $27 \; mmHg$ while it is measured to $26.8 \; mmHg$ in \cite{weibel}. 

\subsection{Equations}

\begin{figure}[h!]
\centering
\includegraphics[height=4cm]{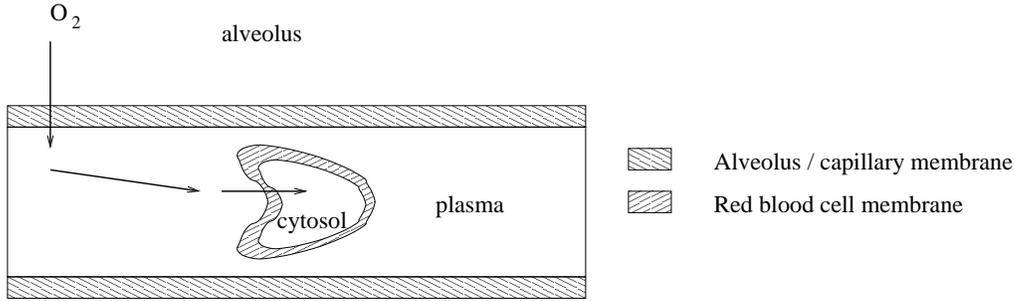}
\caption{Path of oxygen in a pulmonary capillary: oxygen first crosses the alveolus/capillary membrane dissolving into the plasma. Then it is transported in the plasma by convection/diffusion and reaches the red blood cell. It then crosses red blood cell membrane and diffuses inside the cytosol until it reacts with an hemoglobin molecule.}
\label{O2path}
\end{figure}

The path of oxygen from the aveolus into an hemoglobin molecule is described on figure \ref{O2path}. In order to define the equations, we will note $P$ the plasma region and $G$ the cytosol. We will integrate membranes crossing in the boundary conditions in order to simplify the problem, we will note $M$ the boundary between alveolus and capillary and $F=\partial P \backslash M$. We will call $v$ the plasma velocity which is considered as a data in this part, but which comes from calculations of the preceding sections. We will call $a$ the oxygen partial concentration in plasma and $a_g$ the oxygen concentration in the cytosol $G$.

All parameters used in the following equations comes from the references \cite{czer,diff1,diff2,weibel}.

Oxygen equations on $P$ are convection/diffusion equations written in the deformed frame for all $t>0$:

\begin{equation}
\left\{
\begin{array}{ll}
\frac{\partial a}{\partial t} + [\big(v-\frac{d m_c}{dt}\big).\nabla]a - D_1 \bigtriangleup a = 0 & \text{ on $P$}\\
D_1 \frac{\partial a}{\partial n}(Z,t) = \frac{D_2}{\tau_{ac}}(\beta P_{A} - a(Z,t)) \times 1_{\Lambda}(Z+m_c(t).E_c) & \text{ for $Z \in M$ and $t>0$}\\
D_1 \frac{\partial a}{\partial n} = 0 & \text{ on $F$}\\
D_1 \frac{\partial a}{\partial n} = \frac{D_2}{\tau_G}(a_g - a) & \text{ on $\partial G$}
\end{array}
\right.
\end{equation}

where $D_1$ is the diffusion coefficient of oxygen in plasma, $D_2$ the diffusion coefficient of oxygen in the tissues, $\tau_{ac}$ is the alveolus/capillary membrane thickness, $\beta$ the solubility coefficient of oxygen in plasma and $P_A$ the oxygen partial pressure in the aveolus. The supplementary term $1_{\Lambda}(Z+m_c(t).E_c)$ corresponds to the fact that the camera is moving and determines when it will reach a region permeable to oxygen. Thus, the set $\Lambda$ corresponds to the portion of capillary which is permeable to oxygen. Typically, in our simulations : $\Lambda = \{ X | 500 \mu m < X_c < L_c + 500 \mu m \}$ with $L_c$ being the capillary typical length ($1 \; mm$), see section \ref{NRH}. We assume that no diffusive flow crosses the inlet and the outlet. Finally $\tau_G$ is the thickness of the red blood cell membrane and we recall that $a_g$ is the concentration of oxygen in the cytosol $G$. 

Note that because $v$ is close to $d m_c/dt$ near the red blood cell, diffusion is the major phenomena involved in the transport of oxygen near the globule. However, close to the wall of capillary, the apparent transport velocity of oxygen is $d m_c / dt \sim 1 \; mm/s$ (because $v \sim 0$) and this velocity is of the same order than the diffusive velocity $D_1/d_c = 24.10^{-10} / 2.10^{-6} \sim 1 mm/s$ ($d_c$ is the mean distance to be crossed by diffusion). Hence convection and diffusion each play a non negligible role in oxygen repartition near the capillary wall (the P\'eclet number is $1$).

We will now write hemoglobin and oxygen equations in the cytosol $G$. Each hemoglobin molecule $H_i^j$ will diffuse in the cytosol (no transport) and can either:

\begin{itemize}
\item capture an oxygen molecule except if $i=4$, at rate $K_i^j$.
\item free an oxygen molecule except if $i=0$, at rate $k_i^j$.
\item change its configuration, $j \rightarrow 1-j$, at rate $c_i^j$.
\end{itemize}

Hence, doing the balance of sources and sinks of each quantities, we can write, in the deformed frame, the following equations for $H_i^j$ concentration $h_j^j$, for all $j = 1,2$ and $t>0$:

\begin{equation}
\left\{
\begin{array}{ll}
\frac{\partial h_i^j}{\partial t} - D_3 \bigtriangleup h_i^j = - K_i^j a_g h_i^j - c_i^j h_i^j + k_{i+1}^j h_{i+1}^j + c_i^{1-j} h_i^{1-j} & \text{ on $G$, for $i=0$}\\
\frac{\partial h_i^j}{\partial t} - D_3 \bigtriangleup h_i^j = - K_i^j a_g h_i^j - k_i^j h_i^j - c_i^j h_i^j + K_{i-1}^j a_g h_{i-1}^j + k_{i+1}^j h_{i+1}^j + c_i^{1-j} h_i^{1-j} & \text{ on $G$, for $i \neq 0,4$}\\
\frac{\partial h_i^j}{\partial t} - D_3 \bigtriangleup h_i^j = - k_i^j h_i^j - c_i^j h_i^j + K_{i-1}^j a_g h_{i-1}^j + c_i^{1-j} h_i^{1-j} & \text{ on $G$, for $i=4$}\\
D_3 \frac{\partial h_i^j}{\partial n} = 0 & \text{ on $\partial G$, $\forall i = 0,...,4$}
\end{array}
\right.
\end{equation}

$D_3$ is the hemoglobin diffusion coefficient in the cytosol. The boundary condition means that hemoglobin does not cross the red blood cell membrane.

Oxygen diffuses into the cytosol and reacts with the different molecules of hemoglobin. The balance for oxygen in the cytosol can then be written in the deformed frame for all $t>0$:

\begin{equation}
\left\{
\begin{array}{ll}
\frac{\partial a_g}{\partial t} - D_2 \bigtriangleup a_g = - \sum_{i<4,j} K_i^j a_g h_i^j + \sum_{i>0,j} k_i^j h_i^j & \text{ on $G$}\\
D_2 \frac{\partial a_g}{\partial n} = \frac{D_2}{\tau_G}(a - a_g) & \text{ on $\partial G$}
\end{array}
\right.
\end{equation}

We assume that the diffusion coefficient of oxygen in the cytosol is smaller than in plasma, because hemoglobin are big molecules. It has been estimated to be of the same order than in the tissue, see \cite{diff3}.

The initial conditions are chosen in order to correspond to poor oxygenated blood with $40 \; mmHg$ of oxygen partial pressure. Initial hemoglobin state is assumed at equilibrium everywhere in the red blood cell with this oxygen partial pressure.

\section{Oxygen and hemoglobin: first numerical results}
\label{NRH}

We will restrict this preliminary study to two similar cases with pulmonary capillaries which geometries are simple bottlenecks as on figure \ref{compgeom}, but with different sizes: the structure is $2 \; mm$ long and the constricted part, where oxygen exchanges will take place, is $1 \; mm$ long (typical length of a capillary see \cite{weibel}). The larger diameter will be $8.5 \; \mu m$ in both cases. The smaller diameter will be $6.5 \; \mu m$ in the first case and $7.5 \; \mu m$ in the second one. These geometries will be crossed by five red blood cells. We assume that oxygen exchanges with alveolus starts at the beginning of the bottleneck.\\\\

\begin{figure}[h!]
\centering
\includegraphics[height=5cm]{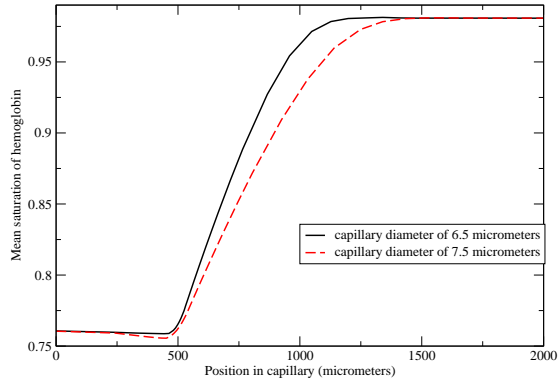}
\caption{Evolution of the mean hemoglobin saturation $SO_2$ in the red blood cells along the capillary. The capillary becomes permeable to oxygen at $500 \mu m$. The larger the radius of the capillary, the slower hemoglobin captures oxygen. Note that because of plasma flow effect on oxygen distribution in the capillary, some slight $SO_2$ inhomogeneities arises before the capillary becomes permeable.}
\label{SO2}
\end{figure}

On figure \ref{SO2}, the mean saturation of oxygen is plotted against the mean position of the red blood cells in the capillary. The first case, with smaller diameter, induces a quicker capture of oxygen by hemoglobin. This is mainly due to the fact that:

\begin{itemize}
\item the distance between the red blood cells and the capillary wall is smaller in the first case than in the second case, hence oxygen crosses the gap quicker when the diameter is small.
\item the red blood cell surface ``available'' for oxygen from capillary wall is larger when the diameter is smaller. Actually, for the larger diameter case, a large part of the surface stays in the middle of the capillary and oxygen needs to cross a larger distance to reach it. 
\end{itemize}

\begin{figure}[h!]
\centering
\includegraphics[height=7cm]{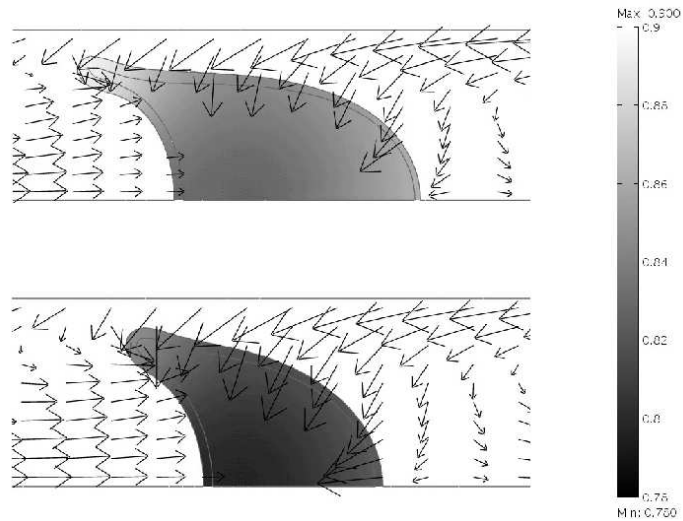}
\caption{Saturation of oxygen in the middle red blood cell. The arrows represent the flow of oxygen in plasma. The upper image corresponds to a $6.5 \; \mu m$ diameter and the lower image corresponds to a $7.5 \; \mu m$ diameter. These images are taken near a $1300 \mu m$ progression in the capillary. Note the differences in oxygen saturation of hemoglobin.}
\label{globo2}
\end{figure}

These observations are shown on figure \ref{globo2}, where the middle red blood cell is drawn for both cases. Both images are taken at the same abscissa, and their difference in term of oxygen saturation is important. As expected the red blood cell regions the more accessible to oxygen (closer to the capillary wall) are first saturated even though hemoglobin is diffusing inside the cytosol. Note that the less saturated region is on the axis, in the rear part of the red blood cell where oxygen flow is the smaller. This effect is due to the screening made in this region by the red blood cell shape to the diffusive process. The screening effect of diffusion can thus play an important role in biological systems. This influence has already been shown in the case of the lung, see \cite{sapo2}. Note that the presence of the other four red blood cells also limits oxygen capture speed through the same phenomena.

\section*{Conclusion}

In this paper, a full modeling chain of red blood cell and oxygen behaviors in a capillary has been built. This gives access to a large panel of data which should be useful in order to better understand oxygen transfer at capillary level. For instance, the fine tuning of some physiological parameters (for example as $V_{O_2 max}$ in the lungs or as blood speed regulation in the vascular system) should be explained by the phenomena involved in capillaries and help to understand the evolution processes that led to such structures. Moreover, pathological behavior (for example paludism or diabetes) could be integrated in this model and could help to understand the consequences of such dysfunctions.

The prospects of this approach is to develop a three dimensional model and to add collisions between red blood cells and with capillary walls. There also exists an important interaction between erythrocytes: they can combine together to form rolls, up to eight cells. These rolls are maintained by chemical forces which can break under high shears. The integration of this behavior is important because of its role in blood circulation. Moreover the restriction to straight capillary could be removed. Finally, mathematical studies of the method should be performed in order to control precisely the errors made during the modeling.\\\\   

{\bf acknowledgement :\\}
The author would like to thank Philippe Dantan from MSC (Paris 7) for useful discussions on blood properties and for his help with Comsol Multiphysics.


\begin{thebibliography}{1}

\bibitem{arslan} Arslan M., Boyce M.C., {\it Constitutive modeling of the finite deformation behavior of membranes possessing a triangulated network microstructure}, J. of App. Mech., vol. 73, pp 536-543, 2006.

\bibitem{canham} Canham P.B., {\it The Minimum Energy of Bending as a Possible Explanation of the Biconcave Shape of the Human Red Blood Cell}, J. theor. Biol., vol. 26, pp. 61-81, 1970.

\bibitem{grenoble} Cottet G.H. and Maitre E., {\it A level-set formulation of immersed boundary methods for fluid-structure interaction problems}, C. R. Acad. Sci., vol. 338, pp. 581-586, 2004.

\bibitem{czer} Czerlinski G., Levin R. and Ypma T., {\it Hemoglobin/$O_2$ Systems: Using Short-lived Intermediates for Mechanistic Discrimination}, J. Theor. Biol., vol. 199, pp. 25-44, 1999.

\bibitem{dao} Dao M., Lim C.T. and Suresh S., {\it{Mechanics of the Human Red Blood Cell Deformed with Optical Tweezers}}, J. Mech. Phys. Solids, 51, pp. 2259-2280, 2003.

\bibitem{dao2} Dao M., Lim C.T. and Suresh S., {\it{erratum: Mechanics of the Human Red Blood Cell Deformed with Optical Tweezers}}, J. Mech. Phys. Solids, 51, pp. 2259-2280, 2003.

\bibitem{eirich} Eirich F.R., {\it Rheology, theory and applications}, vol. 4, Academic Press, 1967.

\bibitem{sapo2} Felici M., Filoche M., Sapoval B., {\it Diffusional screening in the human pulmonary acinus}, J. Appl. Physiol., vol. 94, pp. 2010, 2003.

\bibitem{felici} Felici M., {\it Physique du transport diffusif de l'oxyg\`ene dans le poumon humain}, PhD thesis, Ecole Polytechnique, 2003.

\bibitem{quar} Fern\'andez M.A., Milisic V., Quarteroni A., {\it Analysis of a geometrical multiscale blood flow model based on the coupling of ODE's and hyperbolic PDE's}, SIAM J. on Multiscale Model. Simul., vol. 4 (1), pp. 215-236, 2005.

\bibitem{diff1} Hsia C.C.W., Chuong C.J.C. and Johnson R.L. Jr, {\it{Red cell distortion and conceptual basis of diffusing capacity estimates: finite element analysis}}, J. Appl. Physiol, vol. 83, pp. 1397-1404, 1997.

\bibitem{diff3} Hsia C.C.W., Johnson R.L. Jr, Shah D., {\it Red cell distribution and the recruitment of pulmonary diffusing capacity}, J. Am. Physio. Soc., vol. 86, pp. 1460-1467, 1999.

\bibitem{jeong} Jeong J.H. et al, {\it Measurement of RBC deformation and velocity in capillaries in vivo}, Microvascular Research, vol. 71, pp 212-217, 2006.

\bibitem{glenor} Lenormand G., Henon S. Richert A., Simeon J. and Gallet F., {\it{Direct measurement of the area expansion and shear moduli of the human red blood
cell membrane skeleton}}, Biophysical J., vol. 81, pp. 43-56, 2001.

\bibitem{mauroy} Mauroy B., {\it Hydrodynamique dans les poumons, relations entre flux et g\'emo\'etries}, PhD thesis, ENS Cachan, 2004. {\it http://www.cmla.ens-cachan.fr/$\sim$mauroy/publis/mauroy\_these.pdf}

\bibitem{mills} Mills J.P.,Qie L., Dao M., Lim C.T. and Suresh S., {\it{Nonlinear Elastic and Viscoelastic Deformation of the Human Red Blood Cell with Optical
Tweezers}}, MCB, vol. 1, no. 3, pp. 169-180, 2004.

\bibitem{diff2} Nabors K.L. et al, {\it{Red blood cell orientation in pulmonary
capillaries and its effect on gas diffusion}}, J. Appl. Physiol., vol. 94, pp. 1634-1640, 2003.

\bibitem{pnas} Noguchi H. and Gompper G., {\it{Shape transitions of fluid vesicles and red blood cells in capillary flows}}, PNAS, vol. 102, no. 40, pp. 1459-14164, 2005.

\bibitem{quemada} Quemada D., {\it Towards a unified model of elasto-thixotropy of biofluids}, Biorheology, vol. 21, pp. 423-436, 1984.

\bibitem{secomb1} Secomb T.W., Hsu R. and Pries A.R., {\it{Motion of red blood cells in a capillary with an endothelial surface layer: effect of flow velocity}}, AM; J. Physiol. Heart Circ. Physiol, vol. 281, pp. 629-636, 2001.

\bibitem{di1} Tsukada K. et al, {\it Direct Measurement of Erythrocyte Deformability in Diabetes Mellitus with a Transparent Microchannel Capillary Model and High-Speed Video Camera System}, Microvascular Research, vol. 61, pp. 231-239, 2001.

\bibitem{uzoi} Uzoigwe C., {\it The human erythrocyte has developed the biconcave disc shape to optimise the flow properties of the blood in the large vessesls}, Medical Hypotheses, vol. 67, pp. 1159-1163, 2006.

\bibitem{weibel} Weibel E.R., {\it{The Pathway For Oxygen}}, Harvard University Press, 1984.

\bibitem{winslow} Winslow A.M., {\it Numerical solution of the quasilinear Poisson equation in a nonuniform triangle mesh},
Journal of Computational Physics, vol. 135(2), pp. 128-138, 1997.

\bibitem{yeoh} Yeoh O.H., {\it Characterization of Elastic Properties of Carbon-Black-Filled Rubber Vulcanizates}, Rubber Chem. Technol., vol. 63, pp. 792-805, 1990.

\end{thebibliography}
\end{document}